\pdfoutput=1
\documentclass[aps,prb,nofootinbib,twocolumn,showpacs,floatfix]{revtex4-1}

\usepackage{bm}
\usepackage{amsfonts}
\usepackage[pdftex]{graphicx}
\usepackage[usenames,dvipsnames]{color}
\usepackage{amssymb}
\usepackage{amsmath}
\usepackage{bbm}
\usepackage{amsthm}
\usepackage{hyperref}
\hypersetup{
        colorlinks=false,
}
\usepackage{times}

\usepackage[english]{babel}
\usepackage{blindtext}

\newcommand{\Renyi}{R\'{e}nyi\ }
\newcommand{\sign}{\ensuremath{\mathrm{\sigma}}}
\newcommand{\conf}{\mathcal{C}}
\newcommand{\av}{\mathrm{abs}}
\newcommand{\expv}[1]{\left\langle #1 \right\rangle}
\newcommand{\aexpv}[1]{\expv{\sign}_{\av}}

\newcommand{\mc}[1]{\ensuremath{\mathcal{#1}}}
\newcommand{\tr}{\text{Tr}}
\newcommand{\abs}[1]{\left| #1 \right|} 
\newcommand{\ket}[1]{\left| #1 \right>} 

\let\oldmarginpar\marginpar
\renewcommand\marginpar[1]{\-\oldmarginpar[\raggedleft\tiny\color{red} #1]%
{\raggedright\tiny #1}}

\newcommand{\citationtitle}[1]{{\color{Gray}\small #1}}

\graphicspath{{Figures/}}

\begin{document}

\title{Entanglement and the fermion sign problem in auxiliary field quantum Monte Carlo simulations}
\date{\today}

\author{Peter Broecker}
\author{Simon Trebst}
\affiliation{Institute for Theoretical Physics, University of Cologne, 50937 Cologne, Germany}


\begin{abstract}
Quantum Monte Carlo simulations of fermions are hampered by the notorious sign problem whose most striking manifestation is an exponential growth of sampling errors with the number of particles. 
With the sign problem known to be 
an NP-hard problem and any generic solution thus highly elusive, the Monte Carlo sampling of interacting many-fermion systems is commonly thought to be restricted to a small class of model systems for which a sign-free basis has been identified.
Here we demonstrate that entanglement measures, in particular the so-called \Renyi entropies, can intrinsically exhibit 
a certain robustness against the sign problem in auxiliary-field quantum Monte Carlo approaches and possibly allow for the identification of global ground-state properties via their scaling behavior even in the presence of a strong sign problem. We corroborate these findings via numerical simulations of fermionic quantum phase transitions of spinless fermions on the honeycomb lattice at and below half-filling.
\end{abstract}



\maketitle


While strongly correlated many-fermion systems exhibit some of the most intriguing collective phenomena such as the formation of high-temperature superconductors \cite{Superconductors}, non-Fermi liquids \cite{NonFermiLiquis}, or Mott insulators with fractionalized excitations \cite{SpinLiquids}, their quantitative microscopic understanding remains one of the grand open challenges of theoretical condensed matter physics. With controlled analytical approaches being scarce, one might naturally turn to numerical analysis tools. However, the workhorse of many-particle simulations for bosonic or spin degrees of freedom -- quantum Monte Carlo sampling techniques -- is plagued by the notorious sign problem for fermionic systems, as realized early on \cite{sign_asymptotics_1982,sign_problem_loh_1990}. In short, the sign problem originates from the natural occurrence of negative statistical weights in the fermionic path integral representation tracing back to the minus signs associated with fermion exchange statistics. Ignoring these negative signs to allow for an interpretation as Boltzmann weights in a Monte Carlo approach leads to an exponential growth of statistical errors, and thus computing time, with the number of particles and inverse temperature.

Ever since the early days of quantum Monte Carlo (QMC) simulations it has remained an open problem whether one can systematically overcome this exponential barrier and recover polynomial scaling behavior in fermionic quantum Monte Carlo simulations. Most often, this question is phrased in terms of the quest to identify a basis transformation that leads to a sign-free basis. The idea here is that the sign-problem is not an intrinsic property of the quantum system, but representation-dependent as (at least) the eigenbasis of a many-fermion system (in which the Hamiltonian is diagonal) will not exhibit a sign problem. However, it has been demonstrated that generically identifying such a sign-free basis for an arbitrary many-fermion system is an NP-hard problem \cite{troyer_computational_2005}, making any generic solution highly elusive.
Specific examples of such basis transformations, applicable to certain restricted families of many-fermion systems, have, however, been identified over the years. This includes the meron cluster representation of the Hubbard model in a limited range of attractive and repulsive parameters \cite{meron_cluster_1999}, the fermion bag formulation of four-fermion field theories \cite{fermion_bag_2012}, and, most recently, a Majorana fermion decomposition for the study of SU$(N)$ models with odd $N$ over a wide range of parameters \cite{majorana_rep_2015}.
The lack of a systematic way to identify such basis transformations and its proven NP-hardness have led to the common perception that, despite its fundamental importance, it is impossible to directly address the fermion sign problem.
Instead, numerical efforts have been poured into the development of sophisticated numerical schemes aimed at pushing the exponential barrier to larger system sizes and smaller temperature scales to possibly explore so-far unaccessible regimes of fermionic many-particle systems, such as the development of continuous-time path integral methods for quantum impurity models now commonly employed in dynamical mean-field calculations of strongly correlated electron systems \cite{gull_2011}. Simultaneously, analytical efforts have aimed at identifying abstract, sign-problem free model systems that allow to capture universal low-energy features of microscopic models with a sign-problem, such as the discussion of the onset of antiferromagnetism in metals \cite{berg_2012}. Finally, with the influx of quantum information theory concepts into the description of condensed matter systems orthogonal algorithmic approaches have been devised such as the variational approximation of many-fermion systems via projected entangled-pair states \cite{PEPS}, which are now on the cusp of outperforming more traditional variational Monte Carlo approaches \cite{PEPS2}.

In this manuscript, we return to the original sign problem and explore an alternative route to overcome it. Our guiding idea is to explore whether extracting {\em global} information about the ground state of a many-fermion system via its entanglement properties -- such as the general classification whether it exhibits gapless modes, conventional or topological order -- possibly comes with a smaller sign problem than the measurement of expectation values of observables aimed at providing a full ground-state characterization such as order parameters or two-point correlation functions. This information theoretical angle on the sign problem is rooted in the observation that in the Monte Carlo calculation of entanglement entropies the sign problem enters, as we will explain below, in an {\em additive} instead of multiplicative way as for conventional observables. We find that for certain Monte Carlo flavors, in particular auxiliary-field quantum Monte Carlo approaches, this can lead to a considerable suppression of the sign problem as we demonstrate for spinless fermion models on the honeycomb lattice at and below half filling.
This observation might point a way to an unbiased probe of global ground-state properties of interacting many-fermion systems even in the presence of a strong sign problem.


\noindent {\it Sign problem.--} 
To start our discussion we provide a brief formal overview of the sign problem. It is rooted in the mapping of the quantum system to an equivalent classical system where, for fermionic systems, one might encounter configurations associated with negative or even complex Boltzmann weights. The latter makes it a priori impossible to interpret these statistical weights as probabilities guiding the sampling of a Markov chain of configurations, which is at the core of every Monte Carlo approach. The sign problem arises if one enforces these statistical weights to be positive (and real) by taking their absolute values and interprets these absolute values as the probabilities guiding the Markov process. This is best illustrated by 
considering an observable $\mathcal{O}$, for which one wants to calculate a Monte Carlo estimate of its expectation value $\expv{\mc{O}}$. For each of the configurations $\conf$ of the Markov chain encountered with a Boltzmann weight $p({\mc C})$, one measures the observable $\mc{O(C)}$ as well as the sign $\sigma{(\conf)} = {\mathrm{sgn}}\,p({\mc C})$ of the Boltzmann weight associated with the configuration. The actual expectation value of the observable $\expv{\mc{O}}$ can then be reconstructed~\cite{sign_asymptotics_1982} as %
\begin{equation}
   \expv{\mc{O}} 
   	= \dfrac{\sum {\mc O(C)} p({\mc C})}{\sum p({\mc C})}
	=  \dfrac{\sum {\mc O(C)} \sigma({\mc C}) |p({\mc C})|}{\sum \sigma({\mc C}) |p({\mc C})|}
   	= \dfrac{\expv{\mc{O} \cdot \sigma}_{\av}}{\expv{\sigma}_{\av}}
   \label{eq:obs_signed}
\end{equation}
from measurements $\expv{\ldots}_{\av}$  in the modified ensemble with absolute statistical weights $|p({\mc C})|$.
However, it remains practically impossible to efficiently do this reconstruction of the actual expectation value, since the average sign ${\expv{\sigma}_{\av}}$ in general decreases exponentially
\begin{equation}
  \expv{\sigma}_{\av} = \exp{(-\beta\,N\,\Delta f)}
  \label{eq:sign_asymptote}
\end{equation}
with the number of particles $N$ and inverse temperature $\beta$. The additional factor $\Delta f = f_{\mathrm{fermion}} - f_{\mathrm{abs}}$ is the difference in the free energy densities of the original fermionic system and the one with absolute weights. In the context of world-line Monte Carlo approaches the latter generically describes a {\em bosonic} system, which one obtains by ignoring all $\pi$-phase contributions arising from fermionic particle exchanges. The resulting exponential growth of statistical errors invalidates the polynomial scaling of the Monte Carlo approach.


\noindent {\it Entanglement entropies.--} 
To access global information about the ground state of a many-fermion system, we consider the entanglement entropy for a bipartition of the system into a subsystem $A$ and its complement. It is now well appreciated that these quantum information measures allow to broadly classify the nature of a quantum many-body system via (subleading) corrections to its predominant boundary scaling law \cite{Eisert2010}, such as the formation of a Fermi surface via (multiplicative) logarithmic corrections \cite{Klich06,Wolf06} or the emergence of topological order via $O(1)$ corrections \cite{KitaevPreskill,LevinWen}.
Our particular focus is on the family of \Renyi entropies with an integer index $n \geq 2$
\begin{equation}
  S_n(A) = \dfrac{1}{1 - n}   \log{ \tr{ \left( \rho_A^n \right) }}  \,,
  \label{eq:renyi}
\end{equation}
which generalize the well-known von Neumann entropy (corresponding to the limit $n\to 1$). 
The primary reason to consider these \Renyi entropies in lieu of the more familiar von Neumann entropy is that the \Renyi entropies can be formulated via the so-called replica technique \cite{Holzhey1994,Calabrese04} as the {\em ratio} of two partition sums. Considering the second \Renyi entropy with $n=2$, i.e. $S_2(A) = - \log{\left( \tr{\left( \rho_A^2\right)}\right)}$, 
we have
\begin{equation}
   S_2(A) = -\log{\dfrac{{Z}[A,2,T]}{{Z}^2}} \equiv -\log{\dfrac{Z_1}{Z_0}} \,,
  \label{eq:renyi_n2_z}
\end{equation}
where $Z_1 = {Z}[A,2,T]$ is the partition sum of the replicated system and $Z_0 = {Z}^2$ is the square of the partition function of the original system.
This representation of the \Renyi entropy via the replica technique
is also at the heart of its numerical calculation in quantum Monte Carlo techniques and has recently been 
adopted to various flavors \cite{NumericalReplicaTrick}, in particular fermionic auxiliary field techniques \cite{FermionReplicaTrick1,FermionReplicaTrick2}.
Here our interest in this replica representation arises when considering a quantum system with a fermion sign problem. For such a system the partition sum $Z$ can be split into a product of a partition sum $Z^{\av}=\sum_\conf \abs{p(\conf)}$, where all weights  $p(\conf)$  of the original partition sum have been taken as their absolute values, and the expectation value of the sign
\begin{equation}
  Z = \sum_\conf p(\conf) = \sum_\conf \sigma(\conf)\abs{p(\conf)} = Z^{\av} \cdot \expv{\sigma}_{\av}\,.
  \label{eq:split_z_sign}
\end{equation}
Inserting this expression in the replica representation of the \Renyi entropy \eqref{eq:renyi_n2_z} one readily obtains
\begin{align}
  S_2(A) &= - \log{\left( \dfrac{ Z_{1}^{\av}}{Z_{0}^{\av}}\cdot\dfrac{\expv{\sigma_1}_{\av}}{\expv{\sigma_0}_{\av}}\right)} \nonumber \\
  	      &= -\log{ \dfrac{ Z_{1}^{\av}}{Z_{0}^{\av}}} - \log{\dfrac{\expv{\sigma_1}_{\av}}{\expv{\sigma_0}_{\av}}} \nonumber \\
              &= S_2^{\av}(A) + S_2^\sigma(A) \,.
  \label{eq:ee_sign}
\end{align} 
Thus, the \Renyi entropy for a system with a sign problem separates into two {\em additive} contributions, one coming from the partition sum with absolute weights and one arising from the sign. This additive behavior, which has earlier been discussed in the context of variational Monte Carlo approaches \cite{Grover2011}, should be contrasted to the {\em multiplicative} contribution of the sign to the calculation of conventional observables \eqref{eq:obs_signed}. This additive splitting also raises the immediate question of how contributions to the scaling behavior of the \Renyi entropy $S_2(A)$ are split among the two terms, in particular whether the boundary law scaling or any of its subleading contributions can arise solely from the sign contribution $S_2^\sigma(A)$. Or could it be sufficient to only consider the absolute partition sum and its associated entanglement entropy $S_2^{\av}(A)$, which can be computed from a QMC simulation in a straight-forward manner? While we have no general answer to these questions, we will present in the following numerical evidence from auxiliary-field QMC calculations that shows that for some model systems the entanglement entropy of the absolute partition sum can indeed produce the expected scaling behavior of the \Renyi entropy including subleading terms.


\begin{figure}
  \centering
  \includegraphics{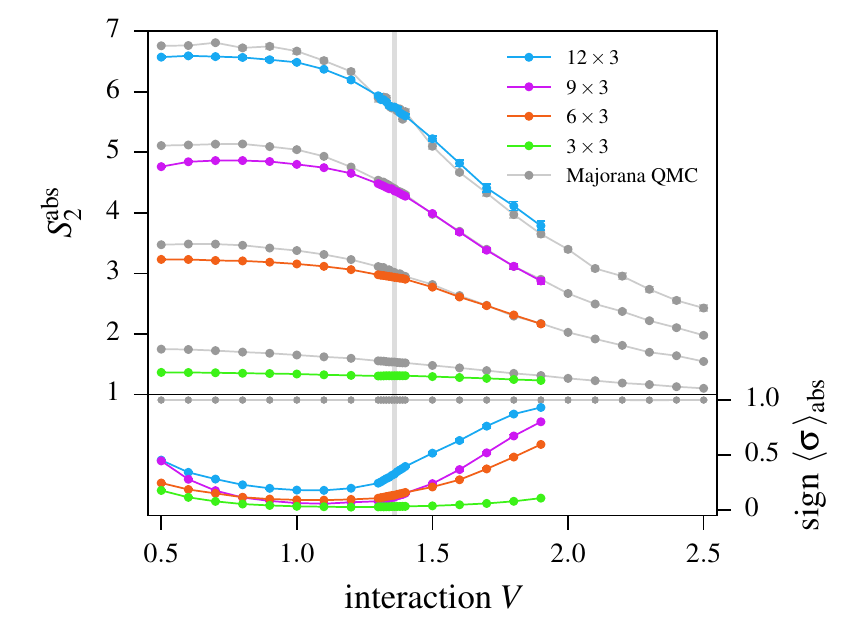}
  \caption{(Color online) 
  		Upper panel: Sign-ignoring entanglement entropy $S_2^{\rm abs}(L)$ for spinless fermions on the honeycomb lattice 
  		at half-filling subject to a nearest-neighbor repulsive interaction $V$. The latter drives a transition from 
		a Dirac semimetal for $V < V_c \approx 1.356(1)$ (indicated by the vertical grey line) to a charge density wave 
		for $V > V_c$. The sign-ignoring data is from projector auxiliary-field QMC calculations using a real-valued 
		Hubbard-Stratonovich transformation at $\theta=40$. 
                Sign-free reference data from Majorana-QMC simulations at $\theta=20$  is indicated
		by the grey dots. 
		Lower panel: The average sign $\expv{\sigma}_{\av}$ indicating the strength of the sign-problem.
   	 	\label{fig:spinless_half_filling_all_Ls} }
\end{figure}

\noindent {\it Spinless Dirac fermions.--} 
As a model system we consider spinless fermions on the honeycomb lattice subject to a nearest-neighbor repulsive interaction $V$ described by a Hamiltonian
\begin{equation}
  H = -t\sum\limits_{\langle i, j \rangle} \left( c_i^\dagger c^{\phantom\dagger}_j + c_j^\dagger c^{\phantom\dagger}_i \right) +  V\sum\limits_{\langle i, j \rangle} n_i n_j \,.
  \label{eq:spinless_hamiltonian}
\end{equation}
At half-filling $\langle n \rangle = 1/2$ this model exhibits a fermionic quantum phase transition, likely in the Gross-Neveu universality class \cite{GrossNeveu,Herbut}, from a Dirac semimetal for small repulsion $V$ to a charge density wave state for large $V$, as recently discussed in various numerical works \cite{NumericalGrossNeveu}. What makes this spinless fermion system particularly interesting for us is that it exhibits a severe sign-problem in the complex-fermion basis underlying conventional auxiliary field (or determinental) QMC techniques \cite{DQMC}, which will be our method of choice, 
while the system can also be recast in a Majorana fermion representation without sign problem \cite{majorana_rep_2015}. This allows to benchmark our results for the sign-ignoring entanglement entropies $S_2^{\av}(A)$ obtained for the complex-fermion case with numerically exact data for $S_2(A)$ from the Majorana fermion approach and in particular distill the sign contribution $S_2^\sigma(A)$ to the \Renyi entropy \eqref{eq:ee_sign}.
Such a comparison of the entanglement entropies is shown in Fig.~\ref{fig:spinless_half_filling_all_Ls} for various system sizes where we cut a system of dimension $2 \cdot L \times 3$ into a strip $A$ of dimension $2 \cdot L \times 1$ and its complement (see Fig.~\ref{fig:honeycomb_cuts} in the appendix for an illustration). While the entanglement entropies show almost perfect agreement in the charge density wave for $V > V_c \approx 1.356(1)$ \cite{NumericalGrossNeveu}, there is a noticeable difference for the Dirac semimetal phase (for $V < V_c$). 
This discrepancy partially reflects the strength of the sign problem in the two phases indicated in the lower panel of Fig.~\ref{fig:spinless_half_filling_all_Ls}, with the average sign $\expv{\sigma}_{\av}$ almost dropping to zero in the Dirac semimetal phase (indicating a strong sign problem), while the average sign quickly recovers (approaching unity) beyond the phase transition into the charge density wave. Note, however, that the deviation of the sign-ignoring entanglement entropy $S_2^{\av}(A)$ from the correct entanglement entropy $S_2(A)$ (calculated in the sign-free Majorana representation) remains almost constant with increasing system size.
\begin{figure}
  \centering
  \includegraphics{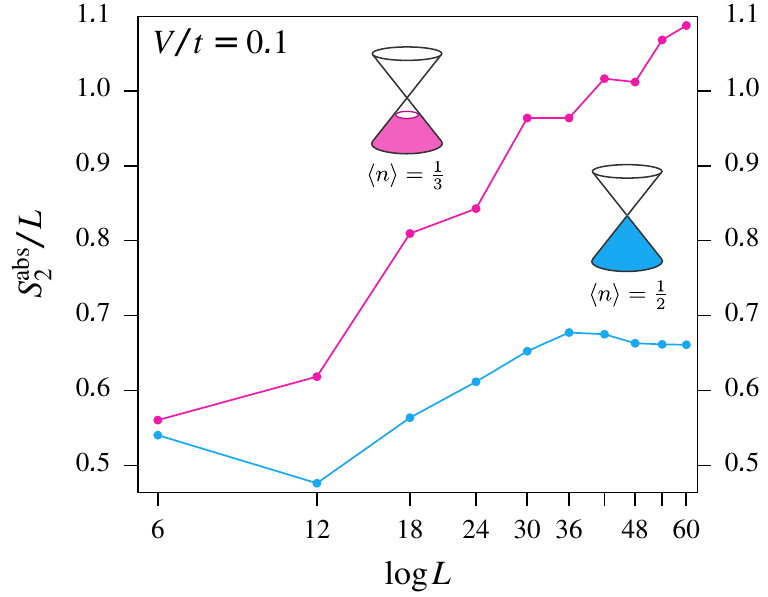}
  \caption{(Color online) Scaling of the sign-ignoring entanglement entropy with system size of a spinless fermion system 
  		on the honeycomb lattice with a small nearest-neighbor repulsive interaction $V/t=0.1$. The system exhibits
		semimetallic states with a Dirac cone at half-filling and a nodal line at one-third filling. 
		Data is from a projector auxiliary-field QMC calculations using a complex-valued Hubbard-Stratonovich transformation at $\theta = 1$,
		for which there is a severe sign problem with $\expv{\sigma}_{\av}\approx0$ for any finite $V$ and all, but the smallest system sizes.
		Error bars are smaller than the symbol sizes.
   		\label{fig:scaling}}
\end{figure}
This immediately raises the question how the expected scaling behavior of the entanglement entropy $S_2(A)$ is split into contributions from the sign-ignoring entanglement entropy $S_2^{\av}(A)$ and the sign entropy $S_2^\sigma(A)$. To probe in particular the scaling behavior of  $S_2^{\av}(A)$ we consider the spinless fermion system of Eq.~\eqref{eq:spinless_hamiltonian} at two different fillings. At half-filling $\langle n \rangle =1/2$ (and small $V$) the system exhibits a Dirac cone with the entanglement entropy expected to follow a boundary law $S_2(L)  = a\ell + \dots$ with the length of the boundary, in our case $\ell = L$ (and $a$ some non-universal perfactor). 
At one-third filling $\langle n \rangle =1/3$ (and small $V$) the partially filled band structure exhibits a Fermi surface with a {\em nodal line} of gapless modes, which leads to a violation of the boundary law with a multiplicative, logarithmic correction arising in the entanglement entropy \cite{Klich06,Wolf06}, i.e. $S_2(L) = a \ell \log{\ell} + \dots$, where $a$ is again some non-universal prefactor.
In Fig.~\ref{fig:scaling} we show numerical results for the sign-ignoring entanglement entropies $S_2^{\av}(A)$ calculated for cuts $A$ of dimension $2 \cdot L \times L/3$ in a system of $2 \cdot L\times L$ sites (see Fig.~\ref{fig:honeycomb_cuts} in the appendix for an illustration). The data in this plot is obtained from auxiliary-field QMC calculations using a {\em complex}-valued Hubbard-Stratonovich transformation, which allows for the calculation of considerably larger system sizes than simulations based on the real-valued Hubbard-Stratonovich transformation -- however, these complex-valued calculations are known \cite{DQMC} to come with a much more severe sign problem, with $\expv{\sigma}_{\av}$ suppressed to zero for any finite $V$ and all but the smallest system sizes.
Remarkably, our numerical results of Fig.~\ref{fig:scaling} suggest that the sign-ignoring entanglement entropies $S_2^{\av}(A)$ completely reflect the scaling behavior of the full entanglement entropy $S_2(A)$, with the data for the half-filled case clearly seen to flatten out for large system sizes indicative of the boundary-law scaling, while the data for the one-third filled case nicely follows the logarithmic scaling behavior expected for the total entanglement entropy $S_2(L)$. This behavior should be contrasted to the results of a recent {\em variational} Monte Carlo study \cite{Grover2011} of model wavefunctions for gapless spin liquids (obtained from a Gutzwiller projection of a Fermi sea), where it was found that the multiplicative logarithmic contribution to the entanglement entropy for a system with gapless nodal lines was solely arising in the sign contribution $S_2^\sigma(A)$.

\begin{figure}
\centering
  \includegraphics{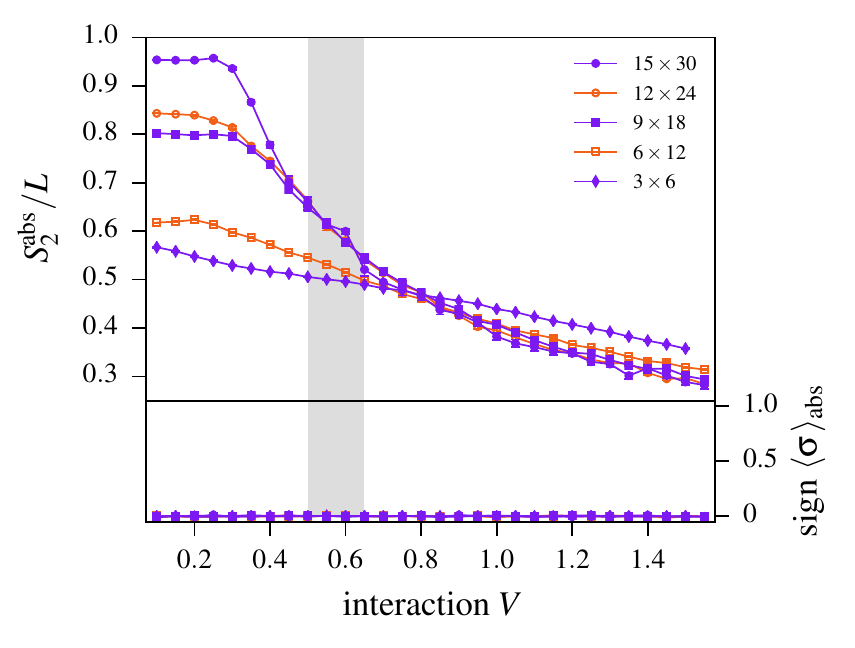}
  \caption{(Color online) 
  		Upper panel: Sign-ignoring entanglement entropy $S_2^{\rm abs}(L)/L$ for spinless fermions on the honeycomb lattice 
  	 	at one-third filling subject to a nearest-neighbor repulsive interaction $V$.
		The numerical data is obtained from projector auxiliary-field QMC calculations using a complex-valued Hubbard-Stratonovich 
		transformation at $\theta=20$, for which there is a severe sign problem for all coupling parameters and system sizes as indicated
		by the vanishing average sign $\expv{\sigma}_{\av}$ shown in the lower panel. 
		\label{fig:transition_doped}}
\end{figure}

Finally, we present results for the quantum phase transition driven by the nearest-neighbor repulsion $V$ in the one-third filled spinless fermion model \eqref{eq:spinless_hamiltonian}. Figure~\ref{fig:transition_doped} shows results for the sign-ignoring entanglement entropy $S_2^{\av}(A)/L$ for system sizes up to $15 \times 30 = 450$ sites (and 150 fermions), 
well beyond the system size limit of any other numerical fermion technique. With increasing $V$ a clear transition in the scaling behavior is seen, going from the logarithmic scaling behavior for the nodal liquid at small $V$ to a clean boundary law for larger $V$ (with all curves collapsing onto each other) indicative of an ordinary charge-ordered state. This allows to estimate the location of this fermionic quantum phase transition to be roughly located at $V_c/t \approx 0.58 \pm 0.04$ as indicated by the vertical 
bar in Fig.~\ref{fig:transition_doped}.

%

%
%
\noindent {\it Discussion.--} 
%
It is important to note that the ignorance of negative signs in the interpretation of statistical weights manifests itself differently 
in various Monte Carlo flavors. In auxiliary-field QMC techniques the weight of a configuration is given by the determinant of 
the free fermion problem in the Hubbard Stratonovich field. It should be noted that at this level the propagation of the fermions
in imaginary time, including all fermionic exchange terms invoked by the Hamiltonian, have been fully taken into account, but
finally condensed into the single number, which is the determinant. Thus, ignoring the sign of the determinant still allows to capture
some of the essential physics of a fermionic system. This should be contrasted to other Monte Carlo flavors, such as world-line
QMC techniques where the ignorance of negative statistical weights readily implies that one considers bosonic instead of 
fermionic exchange statistics throughout and that the actual meaning of coupling terms in a Hamiltonian are fundamentally 
altered, e.g. when effectively considering ferromagnetic couplings in lieu of antiferromagnetic ones.
In a similar spirit, one cannot expect that variational Monte Carlo approaches will be particularly robust when ignoring negative
statistical weights, which readily implies that one fully ignores the  sign structure of the underlying wavefunction (up to the nodal
structure). We believe that this distinction underlies the observation that \Renyi entropies, e.g. for frustrated spin systems, cannot
be calculated in a similarly efficient way for world-line QMC approaches as presented here and is also the source of the stark 
difference in the splitting of entanglement contributions observed for nodal line wavefunctions reported here and in earlier work 
on gapless spin liquids \cite{Grover2011}.

While our numerical results indicate the tantalizing possibility that it might indeed be possible to extract global information about
the ground state of a many-fermion system from auxiliary-field QMC simulations even in the presence of a severe sign problem,
we do want to point out that it will require further thought to understand whether this holds for a restricted class of wavefunctions
only or might apply in more generality. 
From the point  of view of algorithmic complexity theory, it would be interesting to have some guidance into whether the calculation 
of global ground state properties, such as the entanglement entropy, might generally be considered to be in a different complexity class than the calculation of ordinary observables.
We hope that our work will spur further activity to launch a new attack on the long-standing fermion sign problem.


\noindent {\it Acknowledgments.--}
We thank F. Assaad, B. Bauer, T. Grover, and A. Rosch for discussions. 
P.B. acknowledges partial support from the Deutsche Telekom Stiftung 
and the Bonn-Cologne Graduate School of Physics and Astronomy (BCGS).
The numerical simulations were performed on the CHEOPS cluster at RRZK Cologne 
and the JUROPA/JURECA clusters at the Forschungszentrum J\"ulich.



\appendix 
\label{sec:supplemental}
%


\section{Auxiliary field quantum Monte Carlo}
In the following we provide a short synopsis of the main traits of auxiliary-field (or determinental) quantum Monte Carlo approaches, 
for a more detailed exposition of these techniques we refer to the reviews of Ref.~\onlinecite{DQMC}.
Auxiliary-field quantum Monte Carlo techniques come in two distinct variants that allow simulations either at finite temperatures in the grand-canonical ensemble or directly in the ground state, respectively.
In finite temperature simulations, the object of interest is the partition sum
\begin{equation}
Z = \tr{} \rho = \tr{} e^{-\beta H},\nonumber
\end{equation}
while ground-state simulations employ a projective approach on a trial wavefunction $\ket{\psi_T}$ 
\begin{equation}
\ket{\psi} = \lim\limits_{\beta\rightarrow\infty} e^{-\theta H} \ket{\psi_T},\nonumber
\end{equation}
in order to directly access the ground-state properties of a given Hamiltonian.
Note that the inverse temperature $\beta$ and the projection time $\theta$ can be treated on equal footing in the following discussion
(in which we typically refer to the inverse temperature).

\begin{figure}[b]
\includegraphics[width=\columnwidth]{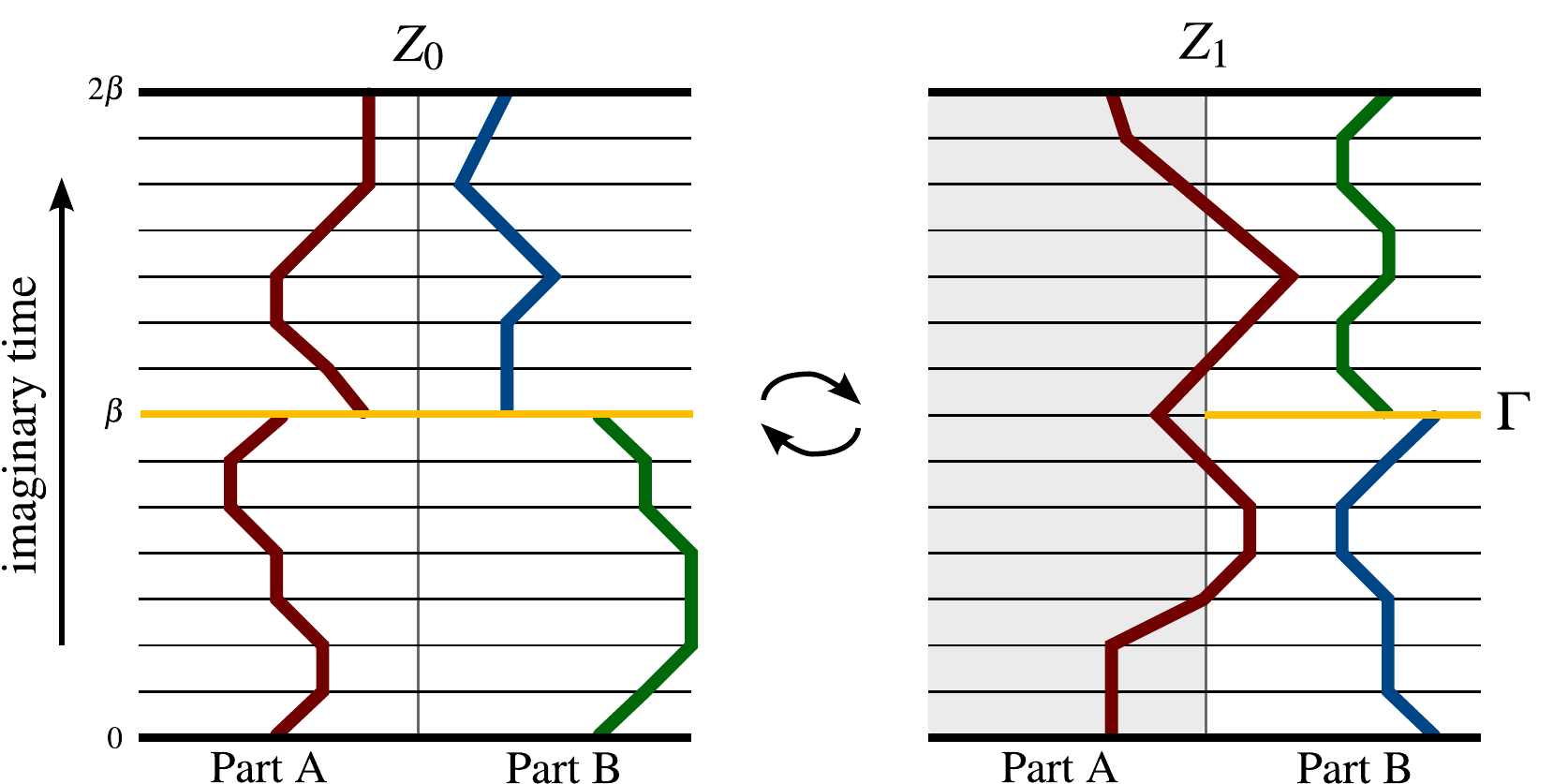}
\caption{(Color online) world-line representation of the two simulation cells used to evaluate $Z_0$ and $Z_1$ respectively.\label{fig:ensemble_switching}}
\end{figure}

In general, neither of the two expressions can be evaluated directly because of quartic terms in the Hamiltonian, such as a density-density interaction.
One typically begins tackling this problem by performing a Trotter-Suzuki decomposition of the inverse temperature into $N$ steps of size $\Delta\tau = \beta / N$
\begin{equation}
  e^{-\beta H} = \prod\limits_{i = 1}^{N} e^{-\Delta\tau H}.\nonumber
\end{equation}
While the quadratic terms of a Hamiltonian can readily be evaluated, the quartic terms cannot be evaluated in a straight-forward manner. 
In auxiliary field QMC techniques one resolves this issue by employing a Hubbard-Stratonovich transformation turning the quartic terms into  quadratic ones at the cost of introducing an additional field -- the Hubbard-Stratonovich field.
Such a decoupling can be performed in various channels. 
For Hubbard models, one may use the magnetization channel, breaking SU(2) symmetry, at the advantage of dealing with real numbers only.
Another possibility that retains the SU(2) symmetry is using the charge channel which leads to complex numbers in the simulation.
The intricacies of the two different decouplings are detailed in Ref.~\onlinecite{DQMC}.
Performing a Hubbard-Stratonovich transformation entails an integration over the Hubbard Stratonovich field, which is performed stochastically within the Monte Carlo procedure.
In the finite temperature algorithm, we will also have to evaluate the trace of the fermion states, which can be performed exactly, since the fermions are free after the decoupling. This results in one determinant per field configuration, hence the alternative name determinantal quantum Monte Carlo.
In the ground state algorithm, the determinant arises as the result of an inner product of the wavefunction with itself. 
More details on the sampling procedure and measurement of observables can be found in the aforementioned literature~\cite{DQMC}.


\section{Entanglement entropies in auxiliary field quantum Monte Carlo}

\begin{figure}
\includegraphics[width=\columnwidth]{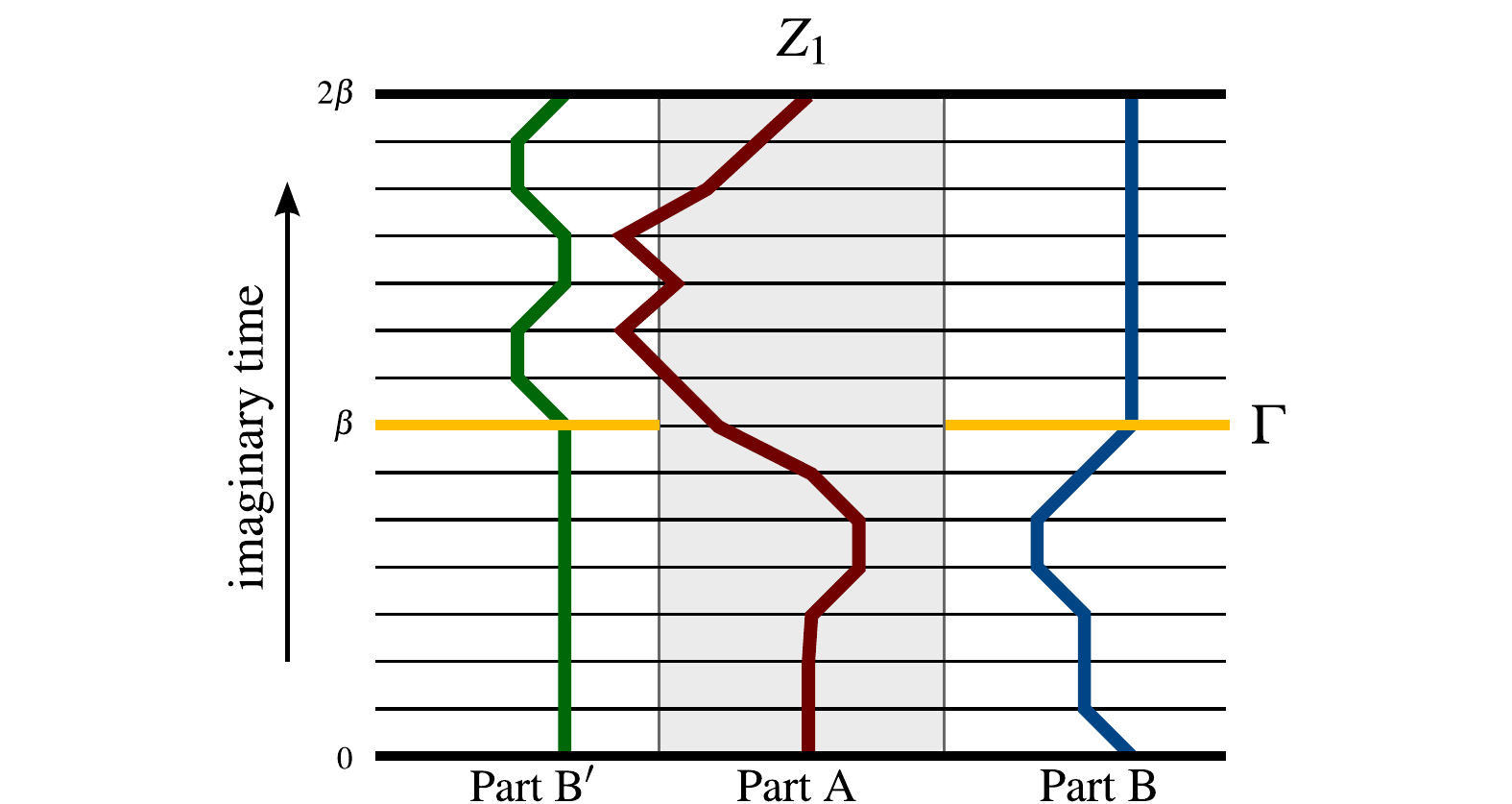}
\caption{(Color online) world-line representation of the modified simulation cell for the calculation of Renyi entropies in DQMC.
\label{fig:modified_renyi_cell}}
\end{figure}

We now turn to the replica trick for the calculation of entanglement entropies, which has been employed in in various analytical and numerical calculation schemes~\cite{Calabrese04, NumericalReplicaTrick, FermionReplicaTrick1,FermionReplicaTrick2}. 
While the concrete implementation in a Monte Carlo simulation depends strongly on the flavor at hand, the core idea is to write a given
\Renyi entropy as the ratio of two partition sums 
\[
   S_2(A) = -\log{\dfrac{{Z}[A,2,T]}{{Z}^2}} \equiv -\log{\dfrac{Z_1}{Z_0}} \,.
\]
Both the numerator and denominator are thus given by a partition sum, albeit with different boundary conditions in the direction of imaginary time. 
A world-line representation of this equation can be found in Fig.~\ref{fig:ensemble_switching}. 
The differences in the boundary conditions correspond to differences in the order of tracing out degrees of freedom and the squaring of the respective density matrix.
For $Z_0$, the entire system is traced out before the square is taken. 
Therefore, none of the degrees of freedom are propagated for times longer than $\beta$ resulting in the original system simply being sampled twice.
The structure of the modified partition sum $Z_1$ is very different because first a partial trace over degrees of freedom in part $B$ is taken before squaring the density matrix and tracing out the remaining degrees of freedom.
This causes the degrees of freedom in part $B$ to be $\beta$-periodic and, like before, appear twice, while those in part $A$ appear only once but are $2\beta$-periodic.

In world-line based Monte Carlo methods, the loop building procedure can easily be adopted to respect these peculiar boundary conditions.
In determinantal QMC, the fermions and their world-lines are traced out and cease to be the central objects. 
Instead, one samples Hubbard Stratonovich field configurations, which makes an adoption of the replica trick somewhat more involved, because the information about the boundary conditions in imaginary time is then solely hidden in the determinant.
One possibility to circumvent this problem is to exploit the fact that the fermions after the  Hubbard Stratonovich transformation are  non-interacting and adopt the correlation matrix method~\cite{FermionReplicaTrick1, CorrelationMatrix} to implement the replica trick. 
This approach, however, turns out to be hampered by large statistical errors and only works well for weakly interacting systems~\cite{FermionReplicaTrick2, FermionReplicaTrick3}.
An alternative approach, applicable also to the strong coupling regime,  was developed by us in an earlier work~\cite{FermionReplicaTrick2}, which demonstrates that one can overcome this problem by directly implementing the replica trick with the help of a modified simulation cell for $Z_1$ as depicted in Fig.~\ref{fig:modified_renyi_cell}.
This idea makes use of the fact, that world-lines from $A$ and $B$ may interact only in restricted intervals of time, i.e. in either one of the intervals $(0, \beta)$ or $(\beta, 2\beta)$.
By ``unfolding'' the simulation cell as depicted in Fig.~\ref{fig:modified_renyi_cell}, we can simulate precisely the same world-line dynamics if we modify the Hamiltonian in such a way that degrees of freedom in the two parts $B$ and $B^\prime$ only propagate during one of the two  intervals in imaginary time, respectively. 
The peculiar boundary conditions in imaginary time have thus been exchanged with a Hamiltonian that depends on imaginary time but that can readily be simulated within the standard DQMC framework, thus allowing to use the replica trick directly.


\section{Convergence properties of entanglement entropy and the sign observable}
\begin{figure}
  \centering
  \includegraphics{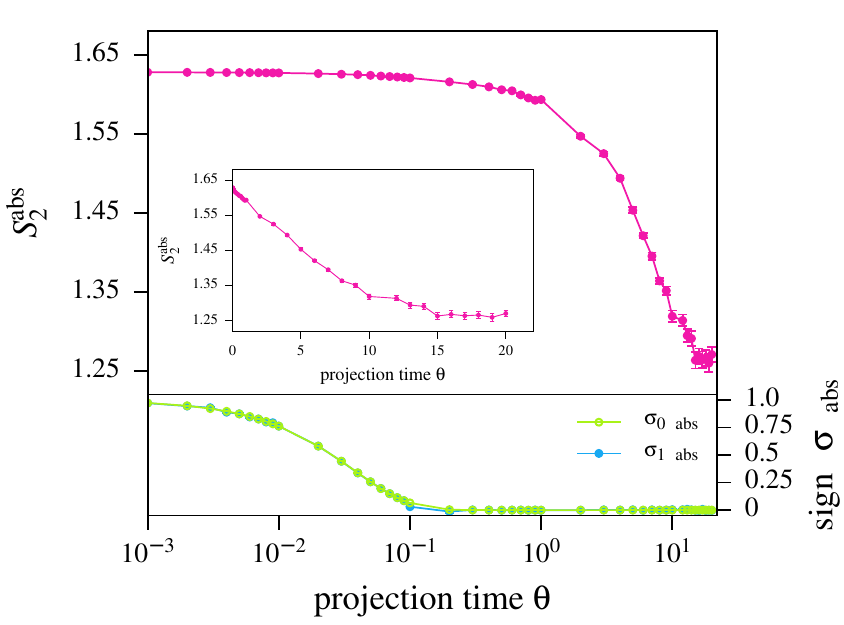}
  \caption{(Color online) Sign-ignoring entanglement entropy $S_2^{\rm abs}(L)$ (upper panel) and expectation values of the two signs $\expv{\sigma_0}$ and $\expv{\sigma_1}$ (lower panel) for spinless fermions on a $2\cdot 3\times 3$ honeycomb lattice at half filling at $V=1.0$. Results were obtained using project auxiliary-field QMC calculations using a complex Hubbard-Stratonovich transformation and plotted against the projection time $\theta$.\ \label{fig:sign_convergence}}
\end{figure}

To demonstrate the severity of the sign problem, we study the convergence of the sign-ignoring entanglement entropy $S_2^{\rm abs}(L)$ and compare it to the expectation value of the signs $\expv{\sigma_0}$ and $\expv{\sigma_1}$ for the two  partition sums sampled in Eq.~\eqref{eq:ee_sign}. In Fig.~\ref{fig:sign_convergence} we show these two observables for a small cluster of spinless fermions on a honeycomb lattice of size $2\cdot 3 \times 3$ at half filling and a nearest-neighbor repulsive interaction of $V=1.0$.
The data reveals several systematic trends that we observe across the entire parameter regime and all system sizes. First, we find that for an extended $\theta$-regime the curves for $\expv{\sigma_0}$ and $\expv{\sigma_1}$ fall on top of each other before posiibly drifting apart for the largest values of $\theta$ (not visible here), which in turn gives rise to a finite sign entanglement entropy $S_2^\sigma$. Second, we find that $\expv{\sigma_0}$ and $\expv{\sigma_1}$ almost vanish well before $S_2^{\rm abs}(L)$ converges. These vanishingly small expectation values  $\expv{\sigma_0}\approx 0$ and $\expv{\sigma_1}\approx 0$ render any direct attempt to determine the sign entanglement entropy $S_2^\sigma$ with a meaningful error bar impossible.


\section{Similarities of entanglement entropies and sign structure}
Finally, we comment on similarities between the entanglement entropy and sign structure of a quantum state. To this end, let us again consider
 the replica formulation of a state, e.g., in its world-line representation in the zero-temperature limit $\beta \to \infty$. In this
limit the imaginary-time boundaries are infinitely far apart and the additional contour $\Gamma$ (see Figs. \ref{fig:ensemble_switching} and \ref{fig:modified_renyi_cell}) arising for the replicated system
can be considered a line defect. As such all ``bulk'' contributions (from intermediate imaginary times) to the entanglement entropies
of the original and replicated systems, i.e. $Z_1$ and $Z_0$ of Eq.~\eqref{eq:renyi_n2_z}, are identical. This gives rise to 
the cancellation of the leading volume contribution in the entanglement entropy scaling \cite{Fradkin} and instead results in the famous
boundary-law \cite{Eisert2010}. A similar argument 
can be made for the expectation values of the signs $\expv{\sigma_1}$ and $\expv{\sigma_0}$ in auxiliary-field QMC 
techniques. While each exchange of two world lines contributes a negative sign to the configuration weight, their overall
number of occurrences in the bulk will be independent of the boundary conditions and the additional contour $\Gamma$. 
As such, we expect that also the sign entropy $S_2^\sigma(A)$ cannot exhibit any extensive scaling behavior. In addition,
both the sign-ignoring entanglement entropy $S_2^{\av}(A)$ (for the absolute value partition sum) and the sign entanglement
entropy $S_2^\sigma(A)$ can be expected to have identical sensitivity to the boundary conditions imposed by the
trace and the additional contour cut $\Gamma$. This leads to the assumption that both $S_2^{\av}(A)$ and $S_2^\sigma(A)$ 
show the same fundamental scaling behavior, i.e. the boundary law and all subleading terms should be manifest in both
entanglement entropies.


%
\begin{figure}[b]
\includegraphics[width=\columnwidth]{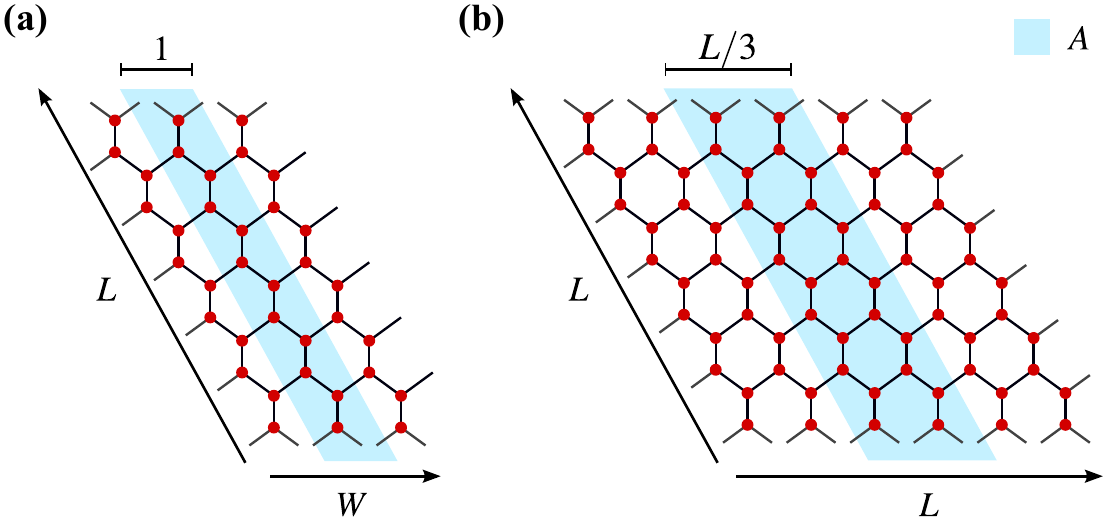}
\caption{(Color online) Geometries of the honeycomb lattice and its bipartitions into subsystems $A$ (shaded in blue) and $B$ 
		used in (a) Fig.~\ref{fig:spinless_half_filling_all_Ls} 
		and (b) Fig.~\ref{fig:scaling}, respectively. 
	\label{fig:honeycomb_cuts}}
\end{figure}

\section{Lattice geometries and cuts}
The lattice geometries and their corresponding cuts are depicted in Fig.~\ref{fig:honeycomb_cuts}.
For the half filled case, it is crucial to capture the Dirac node in the discretized dispersion relation, which requires the width and height of the lattice to be a multiple of $3$.
For the initial comparison of DQMC using a real valued Hubbard Stratonovich transformation and the sign problem free Majorana algorithm shown in Fig.~\ref{fig:spinless_half_filling_all_Ls}, we used a lattice of size $2 \times L \times W$ and set $W = 3$, varying only $L$. 
For Fig.~\ref{fig:scaling}, we were interested in comparing the leading order of contribution to the boundary law. 
We thus chose a lattice of size $2 \times L \times L$, i.e. keeping the aspect ratio constant when increasing the system size. 
The cut is chosen to be free of corners that potentially contribute subleading terms to the area law which we want to avoid for the time being.



\begin{thebibliography}{99}

\bibitem{Superconductors}
J. Zaanen {\em et al.}
\citationtitle{Towards a complete theory of high Tc},
\href{http://dx.doi.org/10.1038/nphys253}{Nature Phys. {\bf 2}, 138 (2006)}.

\bibitem{NonFermiLiquis}
A. J. Schofield,
\citationtitle{Non-Fermi liquids},
\href{http://dx.doi.org/10.1080/001075199181602}{Contemp. Phys. {\bf 40}, 95 (1999)};
H. v. L\"ohneysen, A. Rosch, M. Vojta, and Peter W\"olfle,
\citationtitle{Fermi-liquid instabilities at magnetic quantum phase transitions},
\href{http://dx.doi.org/10.1103/RevModPhys.79.1015}{Rev. Mod. Phys. {\bf 79}, 1015 (2007)}

\bibitem{SpinLiquids}
L. Balents, 
\citationtitle{Spin liquids in frustrated magnets},
\href{http://dx.doi.org/10.1038/nature08917}{Nature {\bf 464}, 199 (2010)}.

\bibitem{sign_asymptotics_1982}
J. E. Hirsch, R. L.  Sugar,  D. J. Scalapino, and R. Blankenbecler, 
\citationtitle{Monte Carlo simulations of one-dimensional fermion systems},
\href{http://link.aps.org/doi/10.1103/PhysRevB.26.5033}{Phys. Rev. B {\bf 26}, 5033, (1982)}.

\bibitem{sign_problem_loh_1990}
E.~Y. Loh, J.~E. Gubernatis, R.~T. Scalettar, S.~R. White, D.~J. Scalapino, and R.~L. Sugar, 
\citationtitle{Sign problem in the numerical simulation of many-electron systems},
\href{http://dx.doi.org/10.1103/PhysRevB.41.9301}{Phys. Rev. B {\bf 41}, 9301 (1990)}.

\bibitem{troyer_computational_2005}
M. Troyer and U.-J. Wiese, 
\citationtitle{Computational Complexity and Fundamental Limitations to Fermionic Quantum Monte Carlo Simulations},
\href{http://dx.doi.org/10.1103/PhysRevLett.94.170201}{Phys. Rev. Lett. {\bf 94}, 170201 (2005)}.

\bibitem{meron_cluster_1999}
S. Chandrasekharan and U.-J. Wiese,
\citationtitle{Meron-Cluster Solution of Fermion Sign Problems},
\href{http://dx.doi.org/10.1103/PhysRevLett.83.3116}{Phys. Rev. Lett. {\bf 83}, 3116 (1999)}.

\bibitem{fermion_bag_2012}
S. Chandrasekharan and A. Li,
\citationtitle{Fermion bag solutions to some sign problems in four-fermion field theories},
\href{http://dx.doi.org/10.1103/PhysRevD.85.091502}{Phys. Rev. D {\bf 85}, (2012)}.

\bibitem{majorana_rep_2015}
Z.-X. Li, Y.-F. Jiang, and H. Yao,
\citationtitle{Solving the fermion sign problem in quantum Monte Carlo simulations by Majorana representation},
\href{http://dx.doi.org/10.1103/PhysRevB.91.241117}{Phys. Rev. B {\bf 91} 241117(R), (2015)}.

\bibitem{gull_2011}
E. Gull, A. J. Millis, A. I. Lichtenstein, A. N. Rubtsov, M. Troyer, and P. Werner,
\citationtitle{Continuous-time Monte Carlo methods for quantum impurity models},
\href{http://dx.doi.org/10.1103/RevModPhys.83.349}{Rev. Mod. Phys. {\bf 83}, 349 (2011)}.

\bibitem{berg_2012}
E. Berg, M.A. Metlitski, and S. Sachdev,
\citationtitle{Sign-Problem-Free Quantum Monte Carlo of the Onset of Antiferromagnetism in Metals},
\href{http://dx.doi.org/10.1126/science.1227769}{Science {\bf 338}, 1606 (2012)}.

\bibitem{PEPS}
P. Corboz, R. Orus, B. Bauer, and G. Vidal,
\citationtitle{Simulation of strongly correlated fermions in two spatial dimensions with fermionic projected entangled-pair states},
\href{http://dx.doi.org/10.1103/PhysRevB.81.165104}{Phys. Rev. B {\bf 81}, 165104 (2010)}.

\bibitem{PEPS2}
P. Corboz, T. M. Rice, and M. Troyer,
\citationtitle{Competing States in the t-J Model: Uniform d-Wave State versus Stripe State},
\href{http://dx.doi.org/10.1103/PhysRevLett.113.046402}{Phys. Rev. Lett. {\bf 113}, 046402 (2014)}.

\bibitem{Eisert2010}
J. Eisert, M. Cramer, and M. B. Plenio, 
\citationtitle{Colloquium: Area laws for the entanglement entropy},
\href{http://journals.aps.org/rmp/abstract/10.1103/RevModPhys.82.277}{Rev. Mod. Phys. {\bf 82}, 277 (2010)}.

\bibitem{Klich06}
D. Gioev and I. Klich,
\citationtitle{Entanglement Entropy of Fermions in Any Dimension and the Widom Conjecture},
\href{http://dx.doi.org/10.1103/PhysRevLett.96.100503}{Phys. Rev. Lett. {\bf 96}, 100503 (2006)}.

\bibitem{Wolf06}
M. M. Wolf,
\citationtitle{Violation of the Entropic Area Law for Fermions},
\href{http://dx.doi.org/10.1103/PhysRevLett.96.010404}{Phys. Rev. Lett. {\bf 96}, 010404 (2006)}.

\bibitem{KitaevPreskill} 
A. Kitaev and J. Preskill, 
\citationtitle{Topological Entanglement Entropy},
\href{http://journals.aps.org/prl/abstract/10.1103/PhysRevLett.96.110404}{Phys. Rev. Lett. {\bf 96}, 110404 (2006)}.

\bibitem{LevinWen} 
M. Levin and X.~G. Wen, 
\citationtitle{Detecting Topological Order in a Ground State Wave Function},
\href{http://journals.aps.org/prl/abstract/10.1103/PhysRevLett.96.110405}{Phys. Rev. Lett. {\bf 96}, 110405 (2006)}.

\bibitem{Holzhey1994}
C. Holzhey, F. Larsen, and F. Wilczek, 
\citationtitle{Geometric and renormalized entropy in conformal field theory},
\href{{http://dx.doi.org/10.1016/0550-3213(94)90402-2}}{Nucl. Phys. {\bf B424}, 443 (1994)}.

\bibitem{Calabrese04}
P. Calabrese and J. Cardy, 
\citationtitle{Entanglement entropy and quantum field theory},
\href{http://iopscience.iop.org/1742-5468/2004/06/P06002/}{J. Stat. Mech. P06002 (2004)}.

\bibitem{NumericalReplicaTrick}
P. B. Buividovich and M. I. Polikarpov, 
\citationtitle{Numerical study of entanglement entropy in SU(2) lattice gauge theory},
\href{http://dx.doi.org/10.1016/j.nuclphysb.2008.04.024}{Nucl. Phys. B {\bf 802}, 458 (2008)};
Y. Nakagawa, A. Nakamura, S. Motoki, and V. I. Zakharov, 
\citationtitle{Entanglement entropy of SU(3) and SU(2) Yang-Mills theories at finite temperature},
\href{http://pos.sissa.it/cgi-bin/reader/contribution.cgi?id=PoS(LAT2009)188}{Proc. Sci. LAT2009, 188 (2009)};
M. B. Hastings, I. Gonzalez, A. B. Kallin, and R. G. Melko, 
\citationtitle{Measuring \Renyi Entanglement Entropy in Quantum Monte Carlo Simulations},
\href{http://journals.aps.org/prl/abstract/10.1103/PhysRevLett.104.157201}{Phys. Rev. Lett. {\bf 104}, 157201 (2010)};
S. Humeniuk and T. Roscilde, 
\citationtitle{Quantum Monte Carlo calculation of entanglement \Renyi entropies for generic quantum systems},
\href{http://journals.aps.org/prb/abstract/10.1103/PhysRevB.86.235116}{Phys. Rev. B {\bf 86}, 235116 (2012)};
N. M. Tubman and J. McMinis,
\citationtitle{Renyi Entanglement Entropy of Molecules: Interaction Effects and Signatures of Bonding},
\href{http://arxiv.org/abs/1204.4731v2}{arXiv:1204.4731};
L. Wang and M. Troyer,
\citationtitle{\Renyi Entanglement Entropy of Interacting Fermions Calculated Using the Continuous-Time Quantum Monte Carlo Method},
\href{http://journals.aps.org/prl/abstract/10.1103/PhysRevLett.113.110401}{Phys. Rev. Lett. {\bf 113}, 110401 (2014)};
C. M. Herdman, S. Inglis, P.-N. Roy, R. G. Melko, and A. Del Maestro,
\citationtitle{Path-integral Monte Carlo method for \Renyi entanglement entropies},
\href{http://dx.doi.org/10.1103/PhysRevE.90.013308}{Phys. Rev. E {\bf 90}, 013308 (2014)};
J. E. Drut and W. J. Porter,
\citationtitle{Hybrid Monte Carlo approach to the entanglement entropy of interacting fermions},
\href{http://dx.doi.org/10.1103/PhysRevB.92.125126}{Phys. Rev. B {\bf 92}, 125126 (2015)}.

\bibitem{FermionReplicaTrick1}
T. Grover, 
\citationtitle{Entanglement of Interacting Fermions in Quantum Monte Carlo Calculations},
\href{http://journals.aps.org/prl/abstract/10.1103/PhysRevLett.111.130402}{Phys. Rev. Lett. {\bf 111}, 130402 (2013)}.

\bibitem{FermionReplicaTrick2}
P. Broecker and S. Trebst,
\citationtitle{R\'enyi entropies of interacting fermions from determinantal quantum Monte Carlo simulations},
\href{http://iopscience.iop.org/1742-5468/2014/8/P08015/}{J. Stat. Mech. P08015 (2014)}.

\bibitem{Grover2011}
Y. Zhang, T. Grover, and A. Vishwanath,
\citationtitle{Entanglement Entropy of Critical Spin Liquids},
\href{http://dx.doi.org/10.1103/PhysRevLett.107.067202}{Phys. Rev. Lett. {\bf 107}, 67202 (2011)}.

\bibitem{GrossNeveu}
D. J. Gross and A. Neveu,
\citationtitle{Dynamical symmetry breaking in asymptotically free field theories},
\href{http://dx.doi.org/10.1103/PhysRevD.10.3235}{Phys. Rev. D {\bf 10}, 3235 (1974)}.

\bibitem{Herbut}
I. F. Herbut,
\citationtitle{Interactions and Phase Transitions on GrapheneÕs Honeycomb Lattice},
\href{http://dx.doi.org/10.1103/PhysRevLett.97.146401}{Phys. Rev. Lett. {\bf 97}, 146401 (2006)};
I. F. Herbut, V. Juricic, and O. Vafek,
\citationtitle{Relativistic Mott criticality in graphene},
\href{http://dx.doi.org/10.1103/PhysRevB.80.075432}{Phys. Rev. B {\bf 80}, 075432 (2009)};
I. F. Herbut, V. Juricic, and B. Roy,
\citationtitle{Theory of interacting electrons on the honeycomb lattice},
\href{http://dx.doi.org/10.1103/PhysRevB.79.085116}{Phys. Rev. B {\bf 79}, 085116 (2009)}.

\bibitem{NumericalGrossNeveu}
L. Wang, P. Corboz, and M. Troyer,
\citationtitle{Fermionic quantum critical point of spinless fermions on a honeycomb lattice},
\href{http://iopscience.iop.org/1367-2630/16/10/103008}{New J. Phys. {\bf 16}, 103008 (2014)};
Z.-Y. Li, Y.-F. Jiang, and H. Yao,
\citationtitle{Fermion-sign-free Majarana-quantum-Monte-Carlo studies of quantum critical phenomena of Dirac fermions in two dimensions},
\href{http://stacks.iop.org/1367-2630/17/i=8/a=085003}{New J. Phys. {\bf 17}, 85003 (2015)};
J. Motruk, A. G. Grushin, F. de Juan, and F. Pollmann,
\citationtitle{Interaction driven phases in the half-filled honeycomb lattice: an infinite density matrix renormalization group study},
\href{http://dx.doi.org/10.1103/PhysRevB.92.085147}{Phys. Rev. B {\bf 92}, 085147 (2015)};
S. Capponi and A. L\"auchli,
\citationtitle{Phase diagram of interacting spinless fermions on the honeycomb lattice: A comprehensive exact diagonalization study},
\href{http://link.aps.org/doi/10.1103/PhysRevB.92.085146}{Phys. Rev. B {\bf 92}, 85146 (2015)}.

\bibitem{DQMC}
For reviews, see e.g. 
F. F. Assaad and H. G. Evertz,
\citationtitle{World-line and Determinantal Quantum Monte Carlo Methods for Spins, Phonons and Electrons},
edited by H. Fehske, R. Schneider, and A. Wei\ss e,
\href{http://link.springer.com/chapter/10.1007/978-3-540-74686-7_10}{Lecture Notes in Physics {\bf 739}, 277 (2008)};
F.~F. Assaad,
\citationtitle{SU(2)-spin Invariant Auxiliary Field Quantum Monte Carlo Algorithm for Hubbard models},
\href{http://www.arXiv.org/abs/cond-mat/9806307}{arXiv:cond-mat/9806307};
R.~R.~dos~Santos, 
\citationtitle{Introduction to quantum Monte Carlo simulations for fermionic systems},
\href{http://dx.doi.org/10.1590/S0103-97332003000100003}{Braz. J. Phys. {\bf 33}, 36 (2003)}.

\bibitem{CorrelationMatrix}
M.-C. Chung and I. Peschel,
\citationtitle{Density-matrix spectra of solvable fermionic systems},
\href{http://dx.doi.org/10.1103/PhysRevB.64.064412}{Phys. Rev. B {\bf 64}, 64412 (2001)};
I. Peschel,
\citationtitle{Calculation of reduced density matrices from correlation functions},
\href{http://dx.doi.org/10.1088/0305-4470/36/14/101}{J. Phys. A: Math. Gen. {\bf 36} L205 (2003)}.

\bibitem{FermionReplicaTrick3}
F. F. Assaad, T. C. Lang, and F. P. Toldin,
\citationtitle{Entanglement spectra of interacting fermions in quantum Monte Carlo simulations},
\href{http://dx.doi.org/10.1103/PhysRevB.89.125121}{Phys. Rev. B {\bf 89}, 125121 (2014)}.

\bibitem{Fradkin}
For a pedagogical discussion of the boundary law, see e.g. E. Fradkin,
\citationtitle{Field Theories of Condensed Matter Physics}, Second Edition, 
Cambridge University Press (2013).

\end{thebibliography}
\end{document}